%% file: aanda-arXiv.tex
\documentclass{aa}  
\usepackage{graphicx}
\usepackage{natbib}
\usepackage[table]{xcolor}
\usepackage{siunitx}
\usepackage{tabularx}
\usepackage{array}
\usepackage{nccmath}
\usepackage{longtable}
\usepackage{lscape}
\usepackage{pdflscape}
\usepackage{txfonts}
\usepackage[colorlinks=true, allcolors=blue]{hyperref}
\usepackage[table]{xcolor}
\usepackage{soul}
\newcommand{\kms}{km~s$^{-1}$}
\newcommand{\ci}{\ion{C}{I}}
\newcommand{\cii}{\ion{C}{II}}
\begin{document}

   \title{[\ci] and [\cii] emission in the circumstellar envelope of IRC +10216}

   \subtitle{I. Observational data and NLTE modeling of the [\ci] emission}

   \author{M. Jeste\thanks{Member of the International Max Planck Research School (IMPRS) for  Astronomy  and  Astrophysics  at  the  Universities  of  Bonn  and Cologne.},
          H. Wiesemeyer,\
          K. M. Menten, and\
          F. Wyrowski
          }

   \institute{Max-Planck-Institut f\"ur Radioastronomie, Auf dem H\"ugel 69, D-53121 Bonn, Germany\\
              \email{mjeste@mpifr.de}
             }

   \date{Received---; accepted---}

% \abstract{}{}{}{}{} 
% 5 {} token are mandatory
 
  \abstract
  {The envelopes of evolved late-type stars on the asymptotic giant branch are characterized by a complex chemistry that near the stellar surface is close to thermochemical equilibrium, but in the outer envelope % owing to the stellar pulsation, and 
  is dominated by radical reactions, assisted by a photo-chemistry driven by the interstellar radiation field.}
  %{The envelopes of evolved late-type stars on the asymptotic giant branch are characterized by a complex chemistry, which close to the star tends to leave equilibrium owing to its pulsation, and which is dominated by \textcolor{red}{radical} reactions in the outer envelope, assisted by a photo-chemistry driven by the interstellar radiation field.}
  {The study at hand aims to describe the distribution of atomic carbon, $\mathrm{C}^0$, throughout the envelope, in support of an improved understanding of its photo-chemistry. Additionally, we also briefly discuss the observation of [\cii] emission towards the star.}
  {We obtain spectra of the [\ci] $\mathrm{^3P_1} \rightarrow \mathrm{^3P_0}$ fine structure line (at 492.160700 GHz) at projected distances of up to $78\arcsec$ from the star. The line profiles are characterized by both direct fitting of Gaussian components, and by modeling the observed line of the [\ci] triplet. We also report the detection of the $\mathrm{^2P_{3/2}} \rightarrow \mathrm{^2P_{1/2}}$ line (at 1900.5369 GHz) from the $\mathrm{C}^+$ fine structure singlet at the central position and at $32\arcsec$ from the star.}
  {The overall picture of the [\ci] emission from IRC~+10216 agrees with more limited previous studies.
  The satisfying agreement between the observed and modeled line profiles, with emission at the systemic velocity appearing beyond one beam ($13\arcsec$ HPBW) from the star, rules out that the $\mathrm{C^0}$ is located in a thin shell. Given that the bond energy of CO falls only 0.1~eV below the ionization threshold of $\mathrm{C}^0$, the absence of observable [\cii] emission from sightlines beyond a projected distance of $\sim 10^{17}$~cm ($\gtrsim 20\arcsec$-- $30\arcsec$) from the star (adopting a distance of 130~pc) does not contradict a scenario where the bulk of $\mathrm{C}^0$ is located between that of CO and $\mathrm{C^+}$, as expected for an external FUV radiation field. This conjecture is also corroborated by a model in which the $\mathrm{C^0}$ shell is located farther outside, failing to reproduce the [\ci] line profiles at intermediate sky-plane distances from the star. Comparing a photo-chemical model adopted from literature with the simplifying assumption of a constant $\mathrm{C}^0$ abundance with respect to the $\mathrm{H}_2$ density 
  (with the $1/r^2$ fall-off of a mass-conserving expansion flow), we constrain the inner boundary of the [\ci] emitting shell, located at $\sim 10^{16}$~cm from the star.}
   {}


\titlerunning{[\ci] and [\cii] emission in the circumstellar envelope of IRC +10216}

\authorrunning{Jeste et al.}
   \maketitle
%-------------------------------------------------------------------
\section{Introduction}

Low- to intermediate-mass stars (0.8 $-$ 8 M\textsubscript{\(\odot\)}) evolve to the asymptotic giant branch (AGB) when they are close to the end of their lives. These objects go through intense mass loss forming a circumstellar envelope (CSE) around them that contains dust and molecules \citep{hoefnerolofsson2018}. Many of the 241 molecules that so far have been identified in astronomical sources, were detected in the CSEs of AGB or red supergiant stars \citep{McGuire2022}, which therefore efficiently enrich the interstellar medium (ISM).
%with various molecules. -> comment out, this is a repetition of the aforesaid.
%The composition of the star during this phase is highly dominated by the surface abundance of carbon to oxygen ratio. 
%-> Not necessarily, we see only what is transported to the surface, so, let's rephrase this:
The abundances of carbon and oxygen at the stellar surface reflect the interplay of nucleosynthesis and convection in the stellar interior.
Stars with C/O $>$ 1 are defined as carbon-rich, those with C/O $<$ 1 as oxygen-rich, and those with C/O $\approx$ 1 are the S-type stars. The abundance of these atoms is mainly due to the third dredge-up, where the material is brought up to the stellar surface and subsequently ejected into the envelope thanks to thermal pulses (TPs) forming in response to instabilities in the helium burning shell. The third dredge-up occurs after each TP (except for a few initial ones, depending on the model,  \citealp[e.g., ][]{Weiss2009}), influencing the C/O ratio in the envelope \citep[further references therein]{Karakas2014}, while the number of TPs depends on the mass loss experienced by the star before it leaves the AGB. %Thus, the depth of the third dredge-up and the mass-loss rate together govern the C/O ratio in the envelope \citep[further references therein]{Karakas2014}.

IRC+10216 (= CW Leonis) is the archetypal carbon-rich AGB star located close to us at a distance between 120 $-$ 140 pc \citep{CrosasandMenten1997,Groenewegen2012} and losing its mass at a high rate of 2 $-$ 4 $\times$ 10$^{-5}$ M$_\odot$ $\text{yr}^{-1} $ \citep{CrosasandMenten1997,DeBeck2018,Fonfria2022}. 

The star has a luminosity of 8600 L$_\odot$ determined from the VLA imaging of the optically thick radio photosphere \citep{Menten2012}, that for a photospheric effective temperature of 2750 K (assumed from SED modeling, \citealp{Menshchikov2001}) entails an optical photospheric diameter of 3.8~AU, while measurements of lunar occultations in the H and K bands estimate a near-infrared diameter of 7.1~AU \citep{Richichi2003}.
%The star has a luminosity of 8600 L$_\odot$ determined by \citet{Menten2012}, %expected from the period-luminosity relation pertaining to TP-AGB stars, and constraining 
%based on the size, 3.8~AU, and % via the Stefan-Boltzmann law if one adopts at 
%photospheric effective temperature, 2750 K, % \citep[further references therein]{Menten2012}, 
%they derive from their imaging with the Very Large Array. 
The star has a very extended molecular envelope moving through the ISM. \citet{Mauron1999} and \citet{Dharmawardena2018} also show that the dust continuum emission extends out to $\sim$ 200\arcsec\ or more from the central star. % in its environment. 
The interphase between the envelope and the ISM is seen in the ultraviolet images taken with the GALEX satellite \citep{Sahai2010}, and at far-infrared wavelengths with the PACS and SPIRE instruments aboard Herschel \citep{Ladjal2010}. Similar interaction between the two environments is also seen in Mira A as the star shows a bow-shock feature in the southward direction, and a tail extending in the north. The star has a space velocity of 130~\kms\ \citep{Martin2007}, larger than what is seen in IRC +10216 \citep[$\sim$ 91~\kms\ in ][]{Sahai2010}. More than 80 molecular species have been detected in the CSE of IRC +10216 so far \citep[see][and many other publications cited therein]{Cernicharo2000,Agundez2014,DeBeck2018,Pardo2022}. These molecules are important probes of the chemical processes at work in the envelope as different molecules at various levels of excitation trace distinct regions in the CSE: parent molecules such as HCN and $\mathrm{C_2H_2}$ form in the inner, hotter part of the envelope (e.g. \citealp[][further references therein]{cernicharo2015,Agundez2020}), whereas the daughter species (i.e., mainly photodissociation products of the former) are found to be present in the outer parts (\citealp[see Fig.~1 in ][]{Li2014} and \citealp{Millar1994,Millar2000,VandeSande2018} for further understanding of the circumstellar chemistry). 

%The distribution of the CO molecule, an important tracer of the history of mass-loss, is quite widespread in the star's CSE: single dish and interferometric observations \textcolor{red}{(with SMA and ALMA)} show almost concentric shells  \citep{cernicharo2015,Guelin2018}. \st{Observations with the IRAM 30m telescope} \textcolor{red}{The former data were taken with the IRAM 30m telescope and they} show that the CO emission is consistently strong up to the photodissociation radius of 180\arcsec, after which there is a sudden drop \citep{cernicharo2015}, owing to its photodissociation yielding atomic carbon which is further ionized to form $\mathrm{C^+}$ \citep{Morris1983,Mamon1988}.
The distribution of the CO molecule, an important tracer of the history of mass loss, is quite widespread in the star's CSE: Observations with the IRAM 30m telescope \citep{cernicharo2015} and with the SMA and ALMA interferometers \citep{Guelin2018} show almost concentric shells, with strong CO emission up to the photodissociation radius of 180\arcsec after which there is a sudden drop (best discernible in the single-dish maps, \citealp{cernicharo2015}) owing to the molecule's photodissociation yielding atomic carbon which is further ionized to form $\mathrm{C^+}$ \citep{Morris1983,Mamon1988,Schoier2001,Groenewegen2017,Saberi2019}.
 
Emission in the two fine-structure lines of neutral atomic carbon, $\mathrm{C^0}$, from the  $\mathrm{^3P_1} \rightarrow \mathrm{^3P_0}$ and $\mathrm{^3P_2} \rightarrow \mathrm{^3P_1}$ transitions near 492 and 809~GHz, respectively, was first detected in molecular clouds by \citet{Phillips1980} and \citet{Jaffe1985}. [\ci] emission\footnote{We denote neutral and ionized carbon as $\mathrm{C}^0$, respectively $\mathrm{C^+}$, while their fine-structure line emission is labeled [\ci] and [\cii], respectively.} is recognized as a valuable tracer for environments characterized by a range of molecular gas fractions occurring in photo-dissociation regions, while either $\mathrm{C^+}$ or CO trace them only partially \citep[further references therein]{Papadopoulos2004}. However, for circumstellar envelopes, very few studies exist of these [\ci] lines. IRC~+10216 has been targeted in the $^3$P$_1$ $-$ $^3$P$_0$ line with the Caltech Submillimeter Observatory (CSO) 10.4 telescope and the James Clerk Maxwell 15 m telescope (JCMT) \citep{Keene1993,Veen1998}. %Under favourable atmospheric transmission, this line is quite interesting to study the neutral regions and can help better understand the chemical evolution of the CO molecule through its photodissociation. \
As suggested by \citet{Keene1993}, another important carrier of carbon in C-rich circumstellar envelopes is acetylene (C$_2$H$_2$), contributing to the shielding against interstellar UV radiation, and acting as precursor molecule for the subsequent photo-chemistry \citep{Santoro2020,Siebert2022}.
%Literature also suggests that the photodissociation of acetylene (C$_2$H$_2$) by external UV interstellar radiation is one of the major contributors to the high [CI] emission in the outer envelope of this star (\textcolor{green}{CITE from keene the three papers?}).

Thanks to the triple bond between carbon and oxygen, CO has a high bond energy, $\mathrm{11\,109.2 \pm 4.1}$~meV \citep{Darwent1970} that is slightly below the ionization energy of $\mathrm{C}^0$, $\mathrm{11\,260.2880 \pm 0.0011}$~meV \citep{Glab2018}. In an environment characterized by internal or external radiation fields and by gas density gradients, the production of free carbon by photodissociation of CO is therefore accompanied by the production of $\mathrm{C}^+$ (of which minor amounts might arise already before, from the photolysis of carbon-bearing species that dissociate easier than CO). Since the ionization threshold of carbon falls significantly below that of hydrogen (13.598~eV; \citet{Jentschura2005}), the $\mathrm{^2P_{3/2}} \rightarrow \mathrm{^2P_{1/2}}$ transition from the $\mathrm{C}^+$ fine-structure singlet originates from both the warm ionized gas phase of the interstellar medium and from cold, neutral and partially molecular gas ($T \sim$10~000~K, respectively $\sim$100~K, \citealp[][with further references therein]{Wolfire2003}). Its emission from diffuse gas is therefore widespread. However, despite its interest for the analysis of the carbon budget in circumstellar environments, detections of the [\cii] line were restricted to post-AGB stars \citep[e.g.,][]{Cerrigone2012,Bujarrabal2016}, before it was found in IRC +10216 \citep{Reach2022}. 

%\textcolor{magenta}{C+ literature here}\\
%\textcolor{red}{read the old 3 CI papers}

%-- Importance of CO in general\\
%-- CO studies done towards IRC+10216\\
%-- CO dissociates to C+ and C \\
%-- Importance of each line i.e. C+ and C\\
%-- Previous work done in these lines in IRC+10216\\
%-- What we do in this paper\\

Given the importance of atomic and ionised carbon in the circumstellar envelope of IRC +10216, we want to constrain the spatial distribution of the atom. In this publication, we discuss the observations we conducted and data reduction methods in Section \ref{obsDR}, followed by the presentation of the obtained spectra and their analysis (section \ref{results}).
%
%of the spectra observed and the radiative transfer code adopted to confirm the observations in Sections \ref{results} and \ref{radtransfer}.
We then discuss the physical constraints we obtained and contextualize them in view of the existing literature %compare it to the literature
in Section \ref{discussion} and conclude our work in Section \ref{conclusion}. 
A more detailed model, including dedicated photo-chemical network and radiative transfer calculations, will be presented in a follow-up study (Wiesemeyer et al., in preparation, hereafter paper II), along with the analysis of the variable [\cii] emission.

%--------------------------------------------------------------------

\section{Observations and data reduction} \label{obsDR}
\subsection{Observations of the [\ci] line}
%Following the work of \citet{Keene1993} and \citet{Veen1998}, 
We performed observations towards IRC+10216 in the $\mathrm{^3P_1} \rightarrow \mathrm{^3P_0}$ fine structure line of atomic carbon with the Atacama Pathfinder EXperiment 12 m submillimeter telescope (APEX)\footnote{This publication is based on data acquired with the Atacama Pathfinder EXperiment (APEX). APEX is a collaboration between the Max-Planck-Institut f\"ur Radioastronomie (MPIfR), the European Southern Observatory, and the Onsala Space Observatory.}. The star was observed under the project ID M-0108.F-9515C-2021 between 2021 November 15 and 30 with the nFLASH460 receiver, a new facility frontend built by the MPIfR for the APEX, while a Fast Fourier Transform Spectrometer \citep[FFTS;][]{klein2006} was used as the backend, providing a spectral resolution of 61~kHz. nFLASH460 is a dual sideband receiver with two polarisations and covers an intermediate frequency (IF) range from 4$-$8 GHz. We tuned the lower sideband (LSB) to the studied [\ci] line, at a frequency of 492.160700 GHz \citep{Haris2017}. The observations took place in good weather with a precipitable water vapour level below 0.6 mm, translating to sightline system temperatures ranging from 400 to 1200~K. We adopt a forward efficiency ($\eta_{\rm f}$) of 0.95 and a main beam efficiency ($\eta_{\rm mb}$) of 0.48\footnote{\url{https://www.apex-telescope.org/telescope/efficiency/}} to calibrate the spectra to the the main-beam brightness temperature scale ($T_{\rm mb}$). At the frequency of 492~GHz, the telescope has a beamwidth of 12.7\arcsec\ (FWHM). The integrations were carried out in a cross along right-ascension and declination offsets, comprising 33 positions, for a total of nine hours equally shared between the target and a reference 1000\arcsec\, away, safely outside of the astrosphere. The observed sky-plane positions are spaced by 6.5\arcsec\ out to 26\arcsec\,from the star, and by 13\arcsec\,beyond, up to a maximum distance of 78\arcsec.  Frequent line pointings were performed on the CO$(4-3)$ line, with the secondary wobbling with a throw of 120\arcsec\, at a frequency of 1.5~Hz. 

%, which has a native channel spacing of xyx kHz. The spectra were smoothed to an appropriate velocity resolution of xyz km~s$^{-1}$ (see Fig. \ref{fig:avg-ci}.)
We used the CLASS software from the GILDAS\footnote{\url{https://www.iram.fr/IRAMFR/GILDAS/}} package \citep{pety2005} to reduce and further analyse the data. The individual steps consist of masking irrelevant parts of the spectra, subtracting linear baseline fits (with baseline noise figures ranging from 38 mK to 120 mK on $T_{\rm mb}$ scale), and averaging adjacent spectral channels, so as to increase the signal-to-noise ratio while preserving an appropriate velocity resolution of 0.2~km~s$^{-1}$, corresponding to five spectral channels in the unprocessed data. Spectra from equivalent positions (or distances from the center) are averaged with $1/\sigma^2$ weighting (i.e., the inverse of the squared baseline noise).

\subsection{Observations of the [\cii] line}
The $\mathrm{^2P_{3/2}} \rightarrow \mathrm{^2P_{1/2}}$ fine-structure line of ionized carbon, [\cii], at 1900.5369~GHz was observed with the upGREAT 7-pixel, dual polarization receiver array for THz spectroscopy \citep{Risacher2018} aboard the {\it Stratospheric Observatory For Infrared Astronomy} (SOFIA). The [\cii] line was tuned to the upper sideband, so as to avoid a contamination by telluric features from the image band of the mixer (operation at THz frequencies precludes sideband separation). The data was acquired on a total of twelve SOFIA flights, spanning 1.4 cycles of the stellar surface pulsation of 630~days \citep{Menten2012}. Two more flights targeting IRC +10216 were conducted in 2021 and 2022. The typical (median) on-sky observing time was one hour (including calibrations and updates of the gyroscope settings), of which typically 40 min were used for integrating towards the target, equally shared between the on-position and the off-positions on either side (chop-and-nod method, with a throw of $2'$). The data was taken on subscan granularity (40 sec including both chop phases) at elevations ranging from $23^\circ$ to $52^\circ$. Median water vapor columns of $10\,\mathrm{\mu m}$ (as referred to zenith) and single-sideband receiver temperatures of 1770 K resulted in a median single-sideband system temperature of 1870~K.

The spectroscopic analysis was performed with Fast Fourier Transform Spectrometers (XFFTS, \citealt{Klein2012}), covering the instantaneous 4~GHz wide bandpass with 16384 channels, resulting in a channel separation of 244~kHz. Data processing followed steps similar to those applied to the [\ci] data; here we remove second-order spectral baselines and average to a spectral resolution of 1.5~\kms, resulting in a baseline noise of 16~mK (rms value, in Rayleigh-Jeans equivalent main-beam brightness temperature). The conversion from backend count rates to forward-beam antenna temperatures was done using calibration loads at ambient and cold temperature (typically around 295 and 100~K, respectively), and adopting a forward efficiency of 0.97. The beam efficiency was calibrated by frequent observations of Mars, and determined to a coupling ranging from $\eta_\mathrm{mb} = 0.65$ to 0.70 (i.e., to the physically possible limit set by the design of the telescope's optics). This interval reflects the inherent calibration uncertainty. At the frequency of the [\cii] line, the main beam of the instrument's Airy pattern has a width of 14\arcsec\ (FWHM), which is merely 10\% wider than the beam at the frequency of the $\mathrm{^3P_1} \rightarrow \mathrm{^3P_0}$ [\ci] transition and thus facilitates the comparison of the line profiles.

\section{Results} \label{results}
%\subsection{[CI] spectra}
\subsection{[\ci] emission}

\begin{figure}[!htbp]
\centering
\includegraphics[trim=2mm 2.7mm 0.5mm 2.5mm,clip=true,width=\columnwidth]{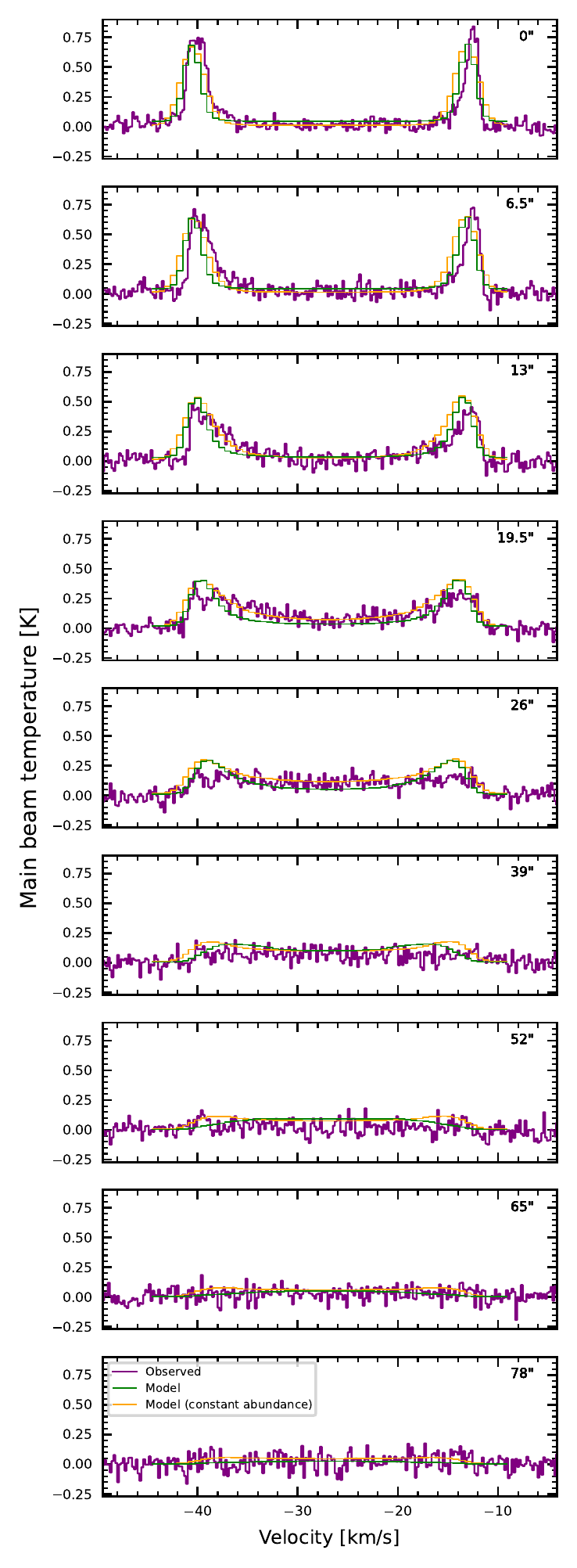}
\caption{[\ci] $^{3}$P$_{1}$~$\rightarrow$ $^{3}$P$_{0}$ averaged spectra towards nine offsets from the star \textit{(in purple)} overlaid with the RATRAN model for variable abundance \textit{(in green)} and constant abundance \textit{(in orange)}.}
\label{fig:avg-ci}
\end{figure}

Since the CO emission from the envelope is largely centro-symmetric \citep{cernicharo2015}, and $\mathrm{C}^0$ and $\mathrm{C}^+$ are products of photochemistry, we average the spectra from the half-beam sampled, equivalent offsets from the star (Fig. ~\ref{fig:avg-ci}). Evidence for deviations from this symmetry will be discussed below; individual spectra, extending to up to $78\arcsec$ from the star, are shown in Fig.~\ref{fig:IndividualPositionsModeled}%s.~\ref{fig:RAneg} to \ref{fig:Decpos}.

\begin{figure}[!htbp]
\centering
\includegraphics[trim=2mm 2.7mm 0.5mm 2.5mm,clip=true,width=\columnwidth]{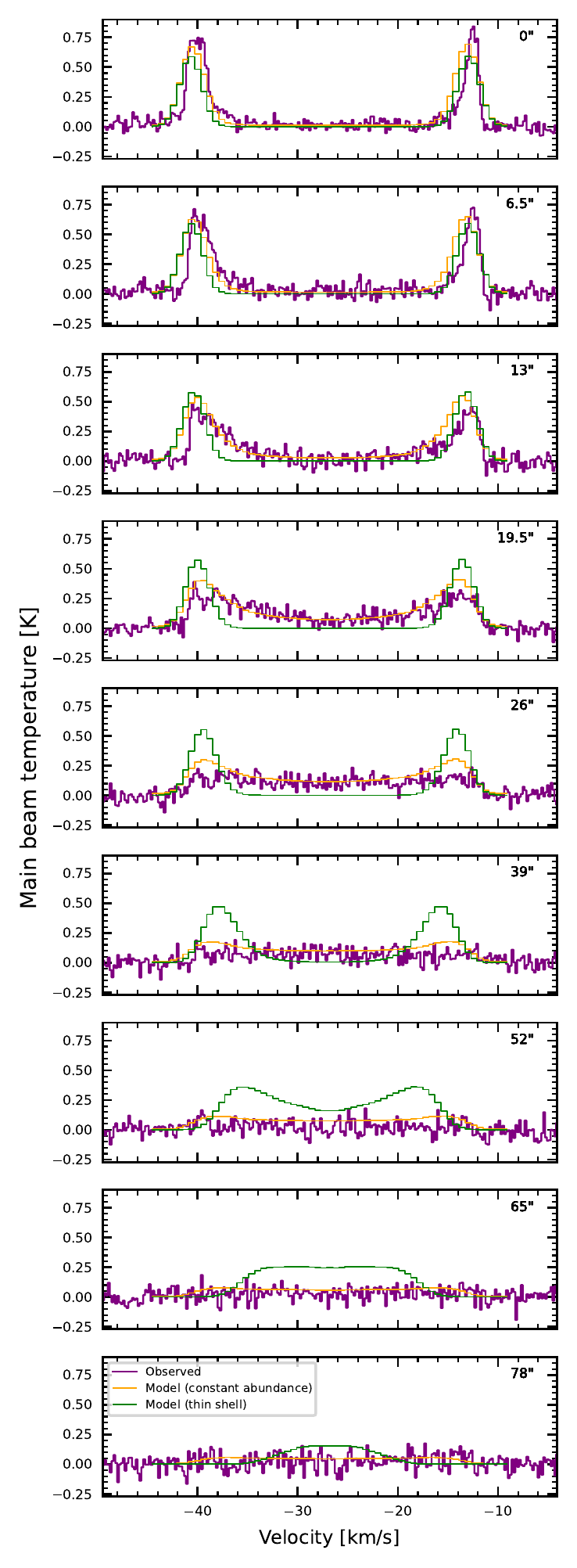}
\caption{As Fig.~\ref{fig:avg-ci}, except that the [\ci] spectra are overlaid with those from the contant-abundance, thick-shell model \textit{(in orange)}, and a thin-shell model \textit{(in green)}. The latter clearly fails on all sightlines except for the two innermost ones.}
\label{fig:avg-ci-thinshell}
\end{figure}

We analyse the spectra using various methods, from a simple phenomenological description assuming constant excitation to a radiative transfer model of the non-LTE excitation of [\ci] throughout a homogeneous envelope, characterized by constant gradients in temperature and density, and expanding at a constant velocity.

\subsubsection{Line profile fitting}

The emission from the center and the innermost sightlines (up to $13\arcsec$ projected distance) displays well separated blue- and red-shifted peaks (horn-shaped line profiles) from the hemispheres expanding towards and away from the observer, respectively. Beyond $19.5\arcsec$ offset, the components start to merge because the sightline-projected expansion velocity decreases. We have fit the observations (up to $26\arcsec$ offsets) with individual Gaussian profiles by masking the emission component at the respective opposite velocity. The resulting parameters are reported in Tab.~\ref{resulttable}. %Starting at $19.5\arcsec$ offset, the components merge, because the sightline-projected expansion velocity decreases, which leads to a partial blend and requires two-component Gaussian fits. %At $39\arcsec$ offset, the line profiles can be described by analytical solutions to the radiative transfer in a circumstellar shell of constant density and excitation, expanding at a constant velocity (\texttt{SHELL} subroutine offered by the CLASS software), which perfectly fits the double-horn profiles.
At $39\arcsec$ significant emission can only be detected after smoothing this spectra to a resolution of 3 \kms\ with a rather flat topped profile of 0.08 K and rms noise of 31.8 mK. For closer inspection we provide the spectrum of the sightline in Fig.~\ref{fig:spec39offset}. At $52\arcsec$ offset and beyond, we see no significant emission anymore. %\textcolor{red}{\st{The fit results are reported in Table~\ref{resulttable}.}} %The mere fact that the latter method is unable to provide a good fit to all spectra demonstrates that the underlying assumptions are inadequate.

The individual spectra (Fig.~\ref{fig:IndividualPositionsModeled}) observed at the same distance from the star show no strikingly remarkable differences. The central sightline features a slightly asymmetric double-peak profile, which can be attributed to a weak absorption in the blue wing, because in the corresponding hemisphere the sightline crosses layers of decreasing excitation. We note, however, that the blue- and red-shifted fractions of the line area display the opposite behaviour. The same holds for the observations of \citet{Keene1993} at $15\arcsec$ (FHWM) resolution, who report peak temperatures of less than 0.4~K, falling 0.3~K below ours. While the discrepancy may be partly due to the individual coupling of the [\ci] emission to the different main beams or due to residual pointing or calibration errors, we note that our line flux of 3.1~K \kms\ is in agreement with that of \citet{Keene1993}. We concede that with the $\simeq$30~years time span between these observations and ours, the [\ci]-emitting shell has expanded by almost 90~AU (or $0.7\arcsec$ at the adopted distance of 130~pc). While such a distance is too short to be noticeable in the beams of single-dish submillimeter telescopes, it may be long enough (17 pulsation periods of 630~d, \citealp{Menten2012}) to possibly alter the line profiles, if the shell and its photochemistry is not in steady-state. From the at most weak asymmetry of the line profile, we conclude that opacity effects cannot be very important in the formation of the line. At offsets of $-19\arcsec$ and $-26\arcsec$ (in either right-ascension or declination), the spectra show more center-filling emission than the direct fits do (i.e., around the systemic velocity of $-26.5$~\kms). %\textcolor{red}{\st{This can be taken as further evidence that the $\mathrm{C^+}$ shell cannot be geometrically thin. Some spatial asymmetry is also entirely conceivable in view of the multiple shells that will be discussed in section 4; as a matter of fact, the direct two-component fits to the spectra at offsets $\pm 19\arcsec$ and $\pm 26\arcsec$ describe the line profiles more closely.}}

\begin{figure}[!htbp]
\centering
\includegraphics[width=\columnwidth]{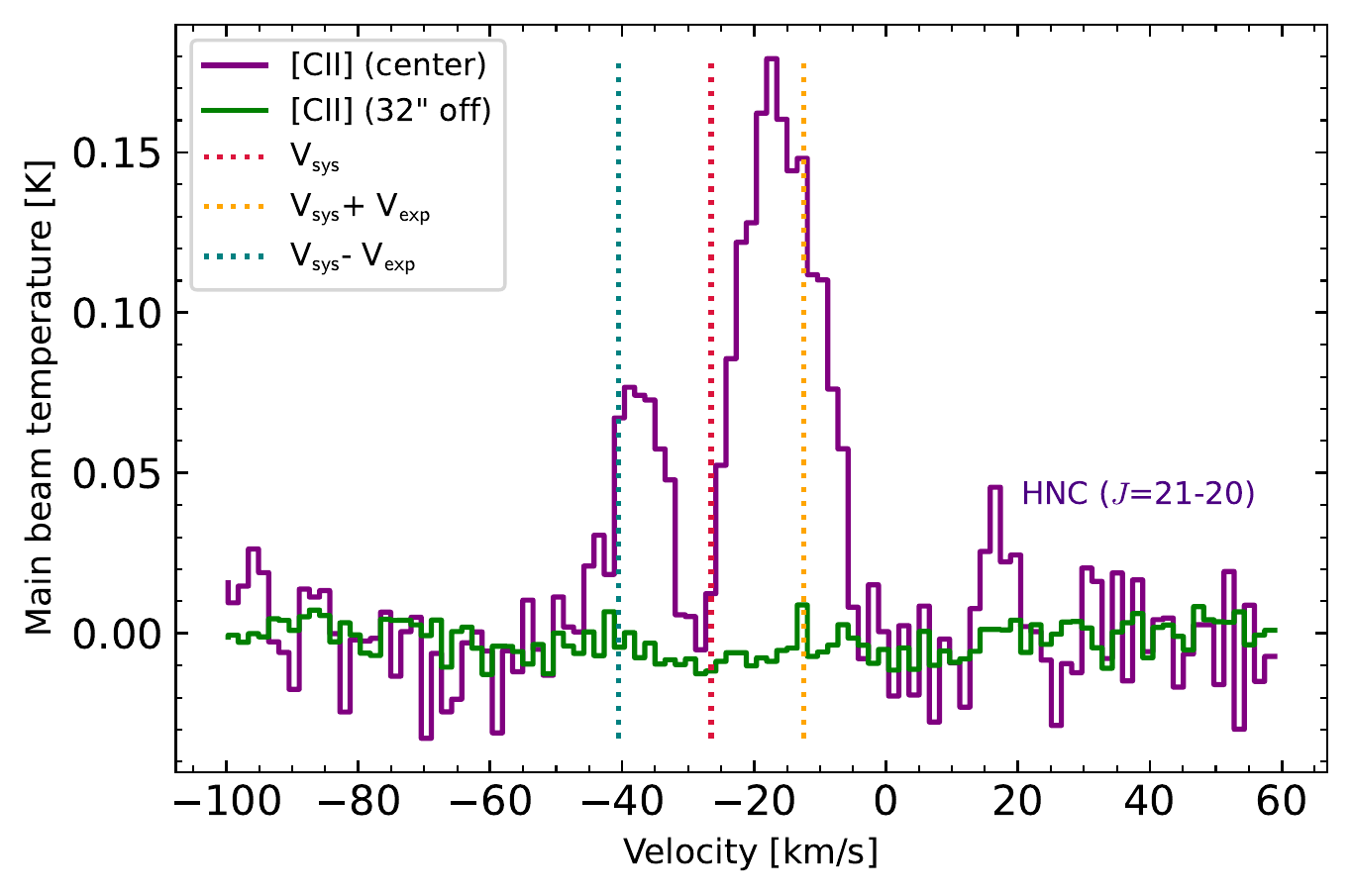}
\caption{[\cii] $\mathrm{^2P_{3/2}} \rightarrow \mathrm{^2P_{1/2}}$ spectrum (2015 - 2021 average) towards the central line-of-sight. The average off-center spectrum (at $32\arcsec$ from the center) is overlaid \textit{in green}. The systemic velocity (V$\mathrm{_{sys}}$) is marked with a red dotted line, and V$\mathrm{_{sys}}$+V$\mathrm{_{exp}}$ and V$\mathrm{_{sys}}$$-$V$\mathrm{_{exp}}$ values are marked with orange and blue dotted line respectively, where V$\mathrm{_{exp}}$ is the gas terminal expansion velocity. The emission feature at 15~\kms\ could be from the HNC $J$ = 21 -- 20 transition, and is marked as such.}
\label{fig:cplus}
\end{figure}

\subsection{$\mathrm{C}^0$ excitation modeling}
Given the ad-hoc approach of the direct fitting, we now use a more sophisticated method to model the excitation of [\ci] throughout an envelope characterized by density, abundance and temperature gradients. We performed radiative transfer modelling to describe the observed spectral lines and determine the abundance and temperature profiles. We used the 1D radiative transfer code RATRAN\footnote{\url{https://personal.sron.nl/~vdtak/ratran/frames.html}} \citep{RATRAN2000}, where a spherically symmetric model is assumed. For the model, we take a distance of 130 pc to the star with a mass-loss rate of 2 $\times$ 10$^{-5}$ M$_\odot$ yr$^{-1}$ \citep{cernicharo2015,Reach2022,Fonfria2022}. Furthermore, the gas is expanding at a velocity of 14 \kms. The line profile function is Gaussian with a 1/e width of 2.0~\kms. Under a turbulent contribution of 1.5~\kms\ \citep[based on observations of CO and C$_2$H,][]{DeBeck2012}, the resulting thermal width of $(\sqrt{2.0^2-1.5^2} = 1.3)$~\kms would be too large to be accounted for by our temperature profile. We note, however, that non-thermal line broadening in circumstellar envelopes is poorly constrained, and that MHD turbulence cannot be ruled out on grounds of CN Zeeman measurements revealing significant deviations from a homogeneous magnetic field structure \citep{Duthu2017}. In such a picture, the fact that the emissions of CO and [\ci] trace spatially distinct envelope regions could then reconcile the different estimates of turbulent line broadening. The excitation model accounts for far-infrared pumping by the dust emission (with assumed gas-to-dust ratio of 100), and assumes that the temperatures of gas and dust are tightly coupled.

\begin{figure*}%[!htbp]
\centering
\includegraphics[width=0.32\textwidth]{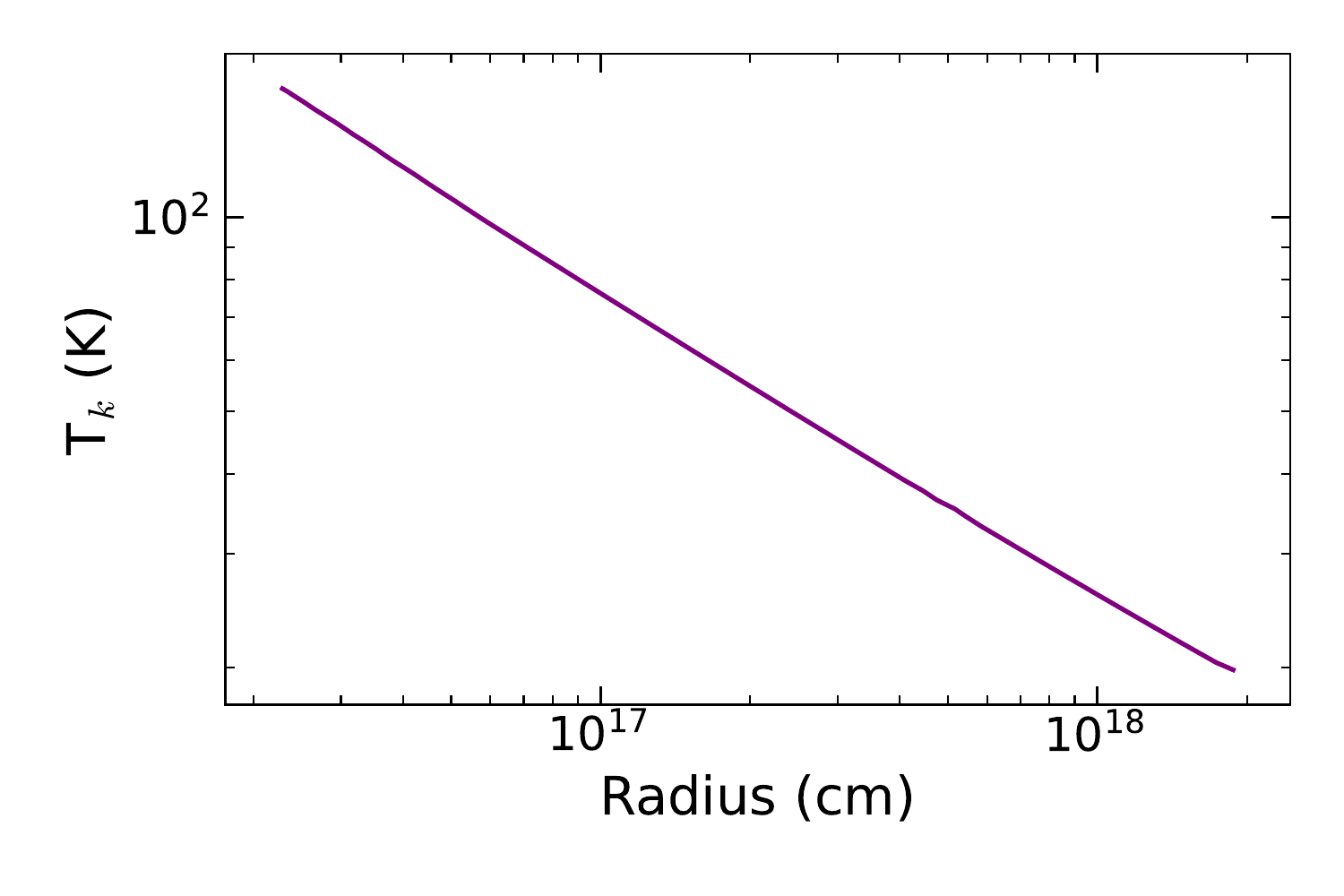}\quad
\includegraphics[width=0.32\textwidth]{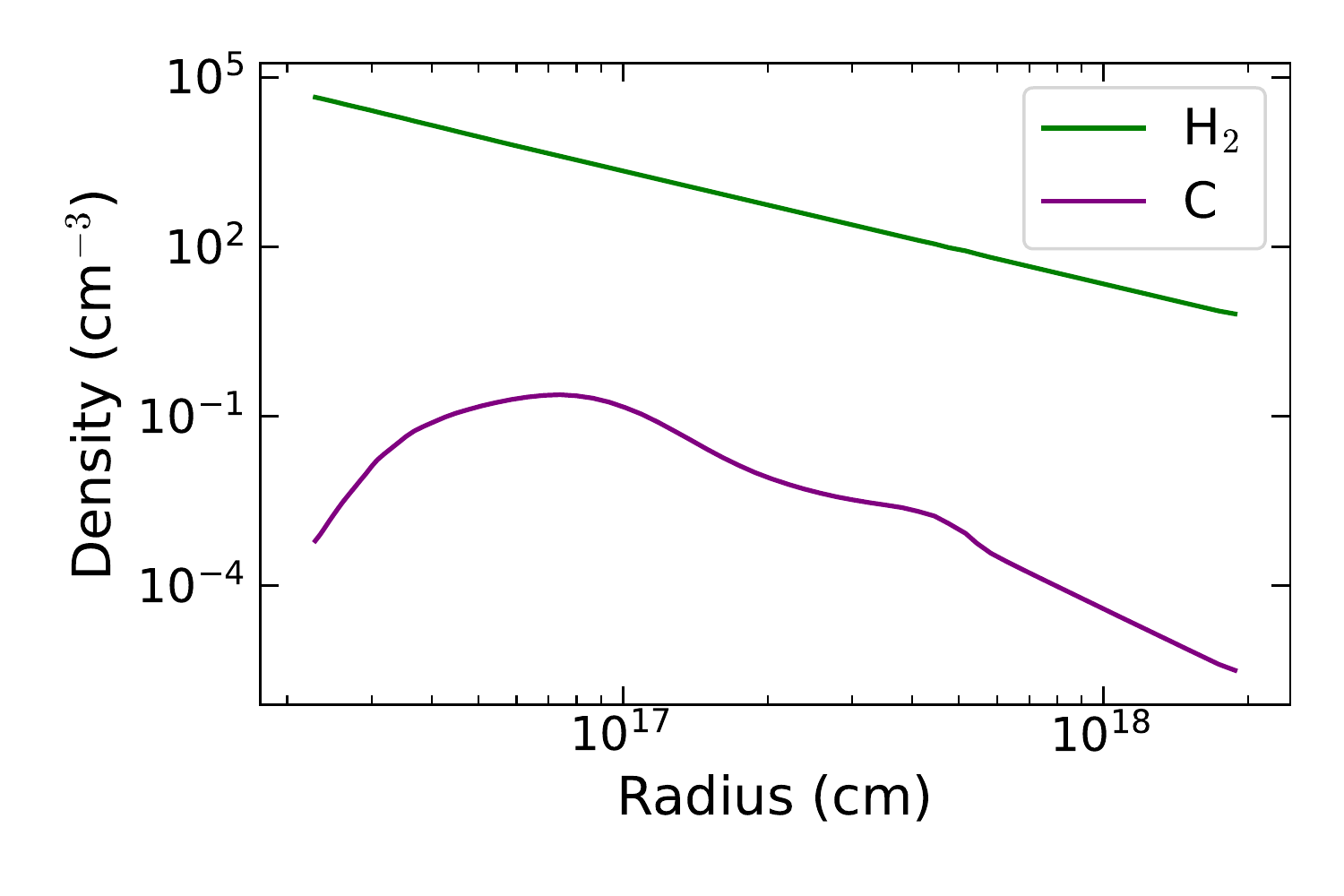}\quad
\includegraphics[width=0.32\textwidth]{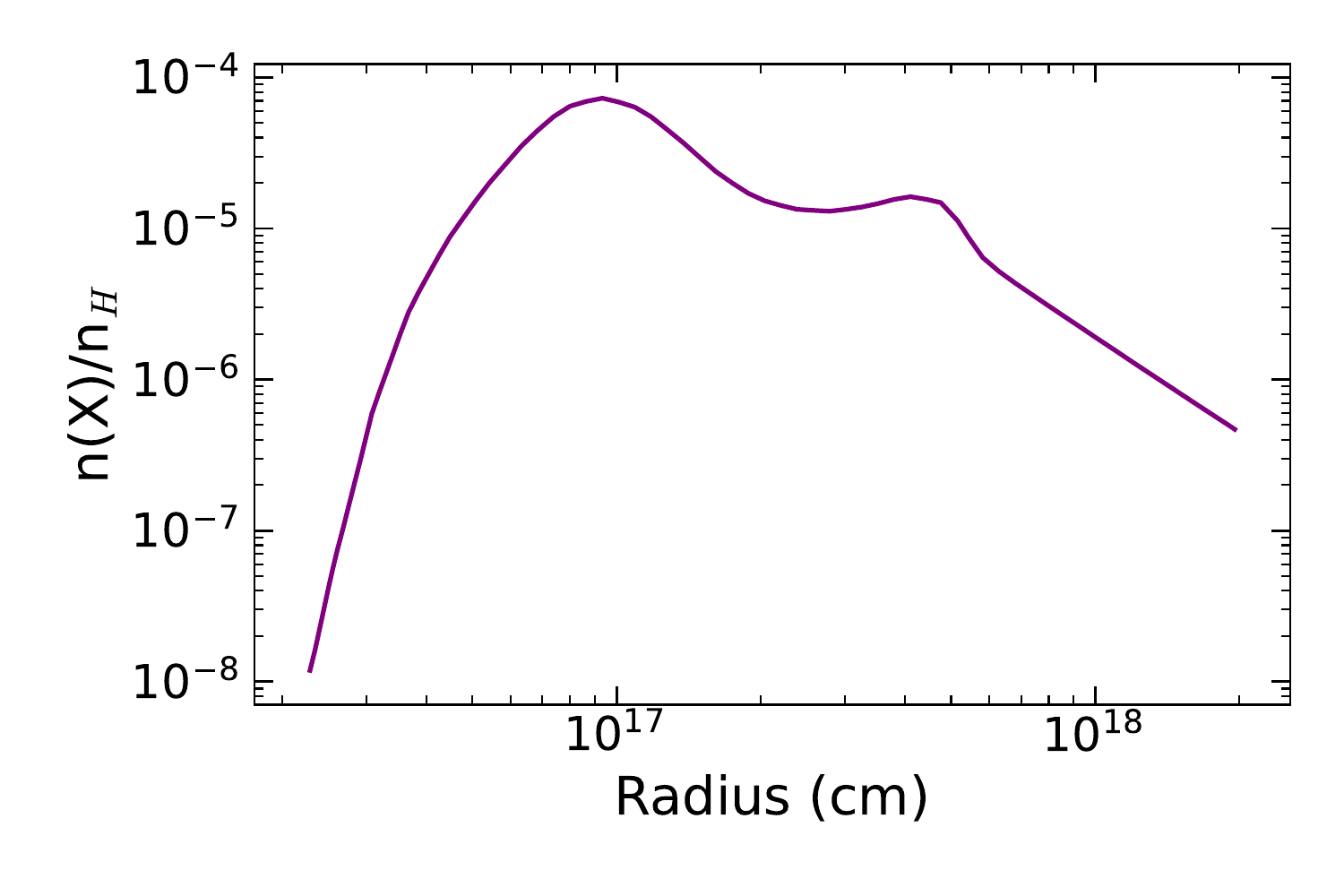}
\caption{Model profiles of \textit{(left:)} kinetic temperature, \textit{(centre:)} densities of $\mathrm{C}^0$ and H$_2$ (the latter taken from \citet{Reach2022}), and \textit{(right:)} $\mathrm{C}^0$ abundance (from \citet{Reach2022}), used as input for the RATRAN model.}
\label{fig:ci-distribution}
\end{figure*}

\citet{Reach2022} assume that the abundances of $\mathrm{C}^{+}$ and $\mathrm{C}^0$ in the outer envelope derive entirely from the photo-dissociation of CO (see their Fig.~3). The abundance of the latter is parameterised by the photo-destruction radius $r_\mathrm{phot}$ and steepness parameter $\alpha$,
\begin{equation}
n(\mathrm{CO})/n(\mathrm{H_2}) = [\mathrm{C}/\mathrm{H}_2]\exp{\left (-\ln{2} (r/r_\mathrm{phot})^\alpha \right )}\,.
\end{equation}
Here we adopt their model with $r_\mathrm{phot} = 3.5 \times 10^{17}$~cm, and $\alpha = 3.0$ \citep[cf.][]{Saberi2019}, for an underlying elemental abundance of $[\mathrm{C}/\mathrm{H}_2] = 8\times 10^{-4}$, of which at the inner boundary of the model 75\% is contained in CO, while the remainder accounts for acetylene ($\mathrm{C}_2\mathrm{H}_2$) and HCN. The model parameters are summarized in Fig.~\ref{fig:ci-distribution} and Table~\ref{inputRATRANmodelparam}. While this model provides a satisfactory description of the azimuthally averaged spectra (Fig.~\ref{fig:avg-ci}, baseline noise 67--128~mK rms) or on individual positions (Fig.~\ref{fig:IndividualPositionsModeled}), we have to test whether it entails uniqueness or not. We therefore overlay the corresponding spectra obtained from adjusting the spectra obtained from a constant-abundance model to the observed ones. A reasonable fit can be obtained with $n(\mathrm{C^0})/n(\mathrm{H}) = 4\times 10^{-5}$ and $r_\text{in} = 4 \times {10^{16}}$~cm, but otherwise the same parameters.This abundance falls an order of magnitude below that of elemental carbon (e.g., $\mathrm{[C/H]} = 2.1\times 10^{-4}$ from young B-type stars, \citealp{Nieva2012}), because it attributes elemental carbon to $\mathrm{C^0}$ at distances from the star where it should still be taken up by mainly CO.

%Interestingly, they seem to fail at different radii: At 13" the constant abundance overestimates the inner velocities, while the emission at the inner velocities is not reproduced by the Reach model at 19.5.", where the constant abundance is doing well. These differences look to me significant. Hence, Reach is apparently doing fine with the inner cut, but high CI abundance seems to extend further than his model.

Fig.~\ref{fig:avg-ci} shows both models overlaid on the observations, reproducing a weakly self-absorbed blue-shifted peak towards the center and the overall distribution of the [\ci] emission. %(except for somewhat higher but still acceptable residuals at $13\arcsec$ from the star). %At $13\arcsec$ offset, the constant abundance model is overestimating the inner velocities. 
On the innermost sightlines, both models reproduce the observed spectral profiles reasonably well. At $19.5\arcsec$ from the star, the variable -abundance model is somewhat underestimating the emission arising closer to the systemic velocity, by 1--2 $\sigma_\mathrm{rms}$, whereas the constant abundance model works well. Beyond this offset, both models agree with each other fairly well. We infer that the model of \citet{Reach2022} can be used to describe the inner boundary of the $\mathrm{C^0}$ shell, while in the constant-abundance model a factor two higher $\mathrm{C^0}$ abundance could be compensated by displacing the inner boundary from $2.3\times 10^{16}$~cm outwards to $4\times 10^{16}$~cm. Such a degeneracy reflects the scaling expected from the adopted density and temperature profiles and the spherical symmetry. Shifting the inner boundary even more towards the photo-destruction radius of CO indeed brings the abundance of $\mathrm{C^0}$ required for a good fit closer to the cosmic, elemental carbon abundance. However, an authoritative chemical model cannot be deduced from these observations alone. We, therefore, continue with a closer look at the [\cii] emission from the central sightline.

\input{RATRANmodel}

\subsection{[\cii] emission}
The average [\cii] $\mathrm{^2P_{3/2}} \rightarrow \mathrm{^2P_{1/2}}$ spectra from the central sightline and at $32\arcsec$ from the star are shown in Fig.~\ref{fig:cplus}. Unlike the [\ci] $\mathrm{^3P_1} \rightarrow \mathrm{^3P_0}$ emission, it is only detected towards the central pixel of the upGREAT array. The outer ring of the array's hexagonal layout displays no detectable [\cii] emission. This finding, and the broader components of the strongly asymmetric double-peak profile (FWHM 8.1 and 13.4~\kms\,in the blue- and red-shifted peak, respectively), less pronounced than that of [\ci], indicate that the emission cannot solely originate from the outer boundary of the circumstellar envelope, exposed to the interstellar radiation field. Additionally, there is an emission feature present at 15 \kms\ velocity, which could be from the HNC $J = 21-20$ transition. This line's small width ($\sim 5$~\kms), which is much smaller than the IRC+10216 CSE's terminal velocity, suggests an origin close to the stellar photosphere, following the conclusions \citet{Patel2009, Patel2011} draw for similarly narrow vibrationally excited %(much) lower-$J$ 
lines 
from various molecules.
Consistent with \citet{Reach2022} we do not detect any emission from the [\ion{$^{13}$C}{ii}] ion. Furthermore, the [\cii] emission was found to be variable on a timescale corresponding to that of the stellar pulsation (paper II). This and the signature of self-absorption in the blue-shifted line component corroborate the presence of a contribution from within the envelope. Although with time the [\cii] line profile shown in Fig.~\ref{fig:cplus} varies both in strength and shape, its overall appearance (broad, strongly asymmetric peaks) is persistent. 

\section{Discussion} \label{discussion}
The [\ci] $\mathrm{^3P_1} \rightarrow \mathrm{^3P_0}$ lines display profiles that are characteristic for optically thin emission from a geometrically extended, expanding envelope. The profiles towards the star, and towards the envelope's limb, could be readily explained by a spatially resolved optically thin shell, whose atomic carbon is produced from the dissociation of CO by the interstellar radiation field: towards the central position, the double-peaked ``horn profile'' is attributed to the front and rear part of the expanding envelope (the slight asymmetry between the blue- and red-shifted peaks will be commented below), while the single-peaked, flat-topped spectra beyond $26\arcsec$ from the star are dominated by sightline elements of which the expansion velocity vector is mostly parallel to the sky plane. The interstitial positions, however, show that the shell cannot be geometrically thin, thanks to the emission close to the systemic velocity filling the double-horn profile. Indeed, a model consisting of a thin, constant-abundance shell extending from $10^{17}$ to $1.6\times 10^{17}$~cm fits only the central spectrum and that at $6.5\arcsec$ distance, but from $13\arcsec$ to $39\arcsec$ distance it cannot provide the emission observed close to systemic velocity (Fig.~\ref{fig:avg-ci-thinshell}). At $52\arcsec$ and beyond, where the shell starts to emit at zero sightline-projected velocity ($-$26.5~\kms in LSR), the double-peak profile is still too pronounced and strong, while the observed emission starts to fall below the detection limit. Likewise, a thin shell extending from $1.2\times 10^{16}$ to $1.6\times 10^{16}$~cm, as advocated by \citet{Reach2022} to explain the [CII] observation of Herschel/HIFI, fails to reproduce our observations as well (see paper II).

Curiously, the [\cii] $\mathrm{^2P_{3/2}} \rightarrow \mathrm{^2P_{1/2}}$ line towards the star (Fig. \ref{fig:cplus}) does not show a narrow double-peak profile either (components $\sim$10~\kms\ wide at corresponding half-maximum), although in a picture where $\mathrm{C}^+$ is produced by the exposure to an external radiation field, the line should originate outwards from the [\ci] emitting layers. On the other hand, producing the [\cii] emission in a layer adjacent to the stellar photosphere is not easily understandable either, because the beam dilution there would require an unusually high emissivity.

\begin{figure}%[!htbp]
\centering
\includegraphics[width=\columnwidth]{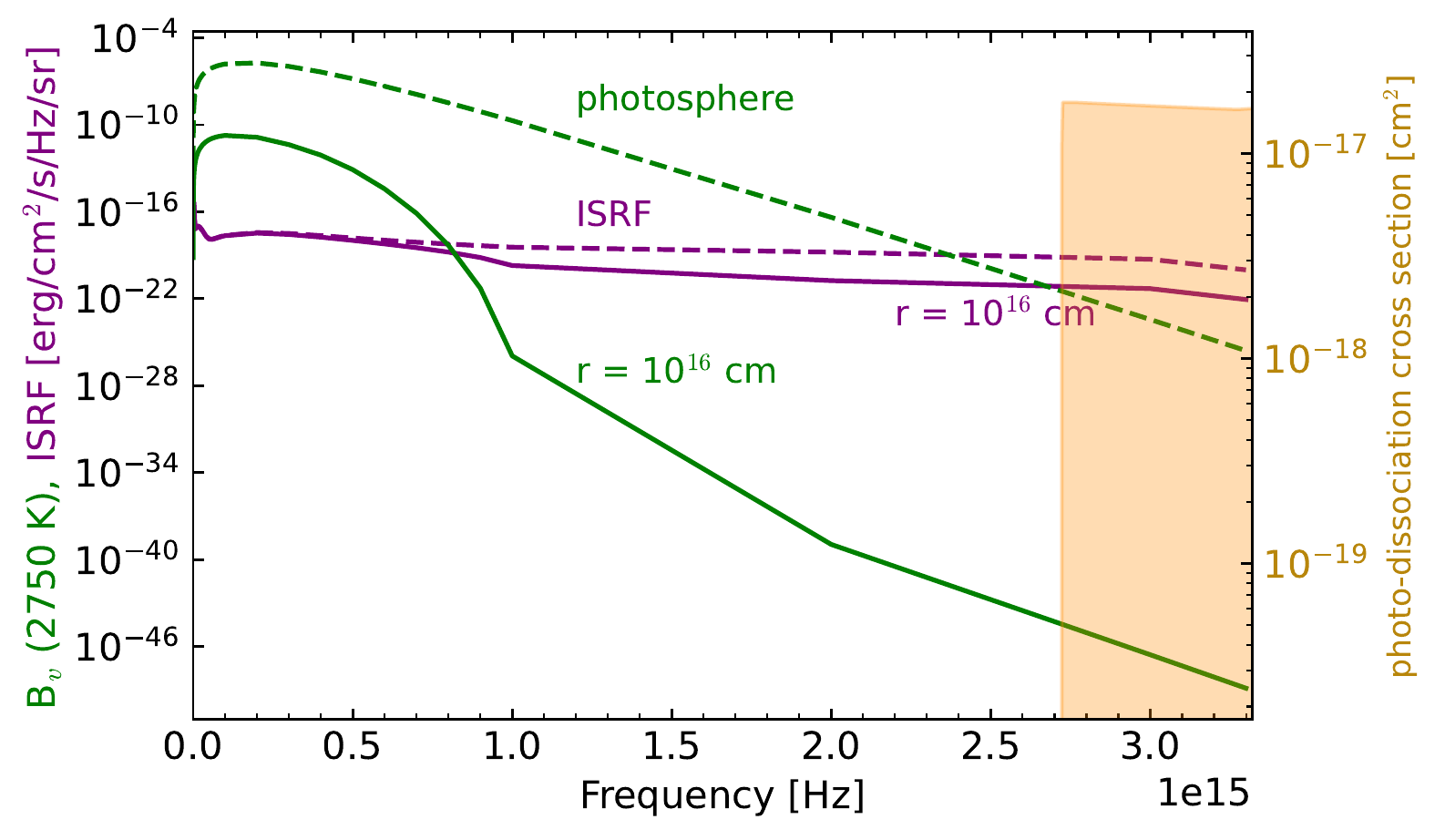}\quad
\includegraphics[width=\columnwidth]{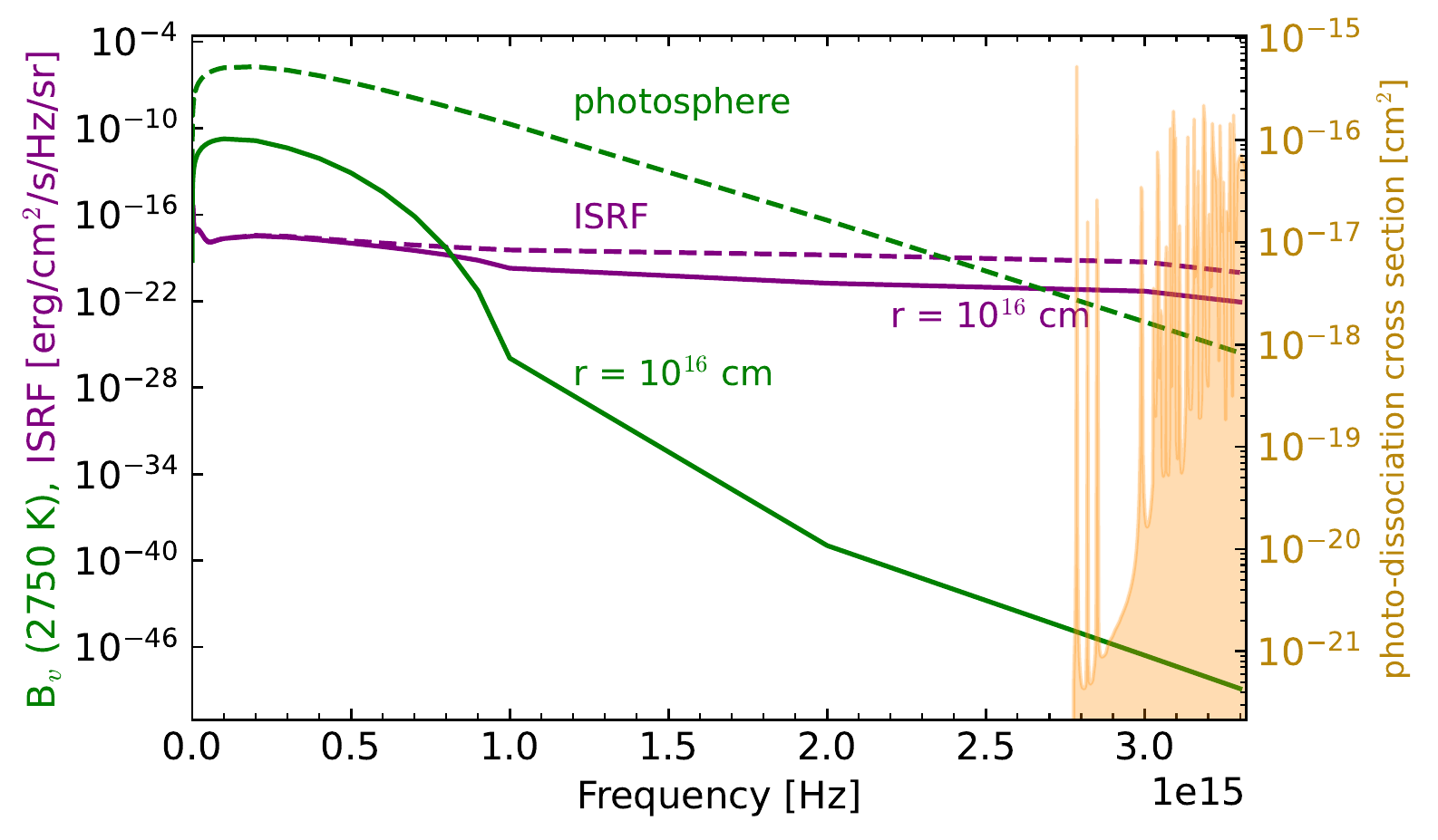}\quad
\caption{Comparison of cross-sections from the Leiden observatory database for photochemistry \citep{Heays2017} for the photo-ionization of \ci~ and the photo-dissociation of CO (gold curves \textit{top} and \textit{bottom}, respectively, with right-hand ordinates) with spectral energy distributions of the attenuated stellar photosphere (green) and the interstellar radiation field (purple) at $10^{16}$~cm from the star (with left-hand ordinate). The corresponding dotted curves refer to the unattenuated and undiluted radiations fields. Dust extinction coefficients are calculated from \cite{Rouleau1991}. The frequency scale is linear, so as to enhance its high-frequency end.}
\label{fig:photo_cross_sections}
\end{figure}

The absence of detectable [\cii] emission at a projected distance of $\ga 20\arcsec$-- $30\arcsec$ from and its presence towards the star (Fig.~\ref{fig:cplus}) suggest that the environment of the inner shell harbours an additional source of ionization. As a matter of fact, at 700~AU distance, the star cannot drive any photo-chemistry ($10^{16}$~cm, Fig.~\ref{fig:photo_cross_sections}). This has led \citet{Reach2022} to the suggestion that mass transfer from the star to a nearby companion locally enhances the UV radiation field, thanks to the shock forming in response to the accretion flow. The presence of such a companion, on a highly eccentric orbit, was indeed advocated by \citet{cernicharo2015}, searching for an explanation for the multiple, partial shells discovered in the optical by light scattering \citep{Mauron1999} and, subsequently, in thermal dust emission \citep[beyond $1\arcmin$ from the star,][]{Decin2011} and CO emission \citep{cernicharo2015,Decin2015,Guelin2018}. The depth and extent of the CO studies demonstrate that a solar-like companion star would provide an explanation for the multiple shells, which form neither in response to thermal pulses, nor to the stellar pulsation, which would result in larger, respectively shorter, spacings. The binary hypothesis is also adopted by \citet[][further references therein]{Siebert2022}, in support of a photo-chemical model of cyanoacetylene ($\mathrm{HC_3N}$) emission from the inner envelope, which would explain the presence of other molecules closer to the star than expected (CH$_3$CN, C$_4$H$_2$) or typical of O-rich stars (H$_2$O). With the data at hand, and the restriction of our models to spherical symmetry, we cannot provide an argument favouring a binary system. We note that \citet{Matthews2018} confirm the  proper motion determined for IRC~+10216's by \citet{Menten2012} with longer time span data and find no evidence of statistically significant astrometric perturbations as would be expected from a binary companion, despite claims to the contrary. As discussed above, some asymmetry seen in the [\ci] profiles is suggestive of a fingerprint of the multiple-shell structure. Given that it shows clearly in CO~$(2-1)$ emission observed at $11\arcsec$ resolution \citep[FWHM, ][]{cernicharo2015}, it should also be discernible in our $13\arcsec$ resolution [\ci] data.  The Fourier transforms analysis of \citet{Guelin2018} reveals a $16\arcsec$ shell spacing in all quadrants except in the south-eastern one, where our direct line-profile fits display larger residuals. Whether this observation is coincidental or not can only be settled by a full, critically sampled [\ci] map.

\section{Conclusion} \label{conclusion}

Following the pioneering observations of atomic carbon in IRC~+10216 by \citet{Keene1993}, we present more sensitive and more extensive observations of the $\mathrm{^3P_1} \rightarrow \mathrm{^3P_0}$ emission from the [\ci] fine structure triplet. Accounting for differences in the data quality, and allowing for an underlying variability which might show after the 30~year time lapse between the observations, we notice an overall good agreement. We present a model employing an abundance gradient that best reproduces the lines profiles at increasing distance from the star -- however, even a model employing a constant $\mathrm{C}^0$ abundance would work. We infer that the inner boundary of the [\ci]-emitting shell is at $\sim 10^{16}$~cm from the star, but can not yet locate it more precisely. The fact that the [\cii] line is seen only towards the inner sightlines through the circumstellar envelope ($\la 4\,000$~AU), while the outer envelope, ranging to up to $\sim$10\,000~AU, is void of [\cii] emission, remains an enigma. Both findings call for a full map of the [\ci] emission, and deeper observations of the [\cii] line. The line profiles obtained so far from both tracers exclude a picture in which atomic carbon is only produced in a thin low-density shell fully exposed to the interstellar radiation field.

\begin{acknowledgements}
This publication is based on data acquired with the Atacama Pathfinder EXperiment (APEX). APEX is a collaboration between the Max-Planck-Institut für Radioastronomie, the European Southern Observatory, and the Onsala Space Observatory. The authors thank the anonymous referee for thoughtful comments. MJ thanks Ankit Rohatgi for making the WebPlotDigitizer tool open source (\url{https://automeris.io/WebPlotDigitizer/}). MJ also thanks Ivalu Barlach Christensen for her help in fixing some python scripts. This research has made use of NASA’s Astrophysics Data System. This work has also made use of Python libraries including NumPy\footnote{\url{https://www.numpy.org/}} \citep{5725236}, SciPy\footnote{\url{https://www.scipy.org/}} \citep{jones2001scipy}, and Matplotlib\footnote{\url{https://matplotlib.org/}} \citep{Hunter:2007}. 
\end{acknowledgements}

\bibliography{references}
\bibliographystyle{aa}

\begin{appendix} 
%\section{\st{Calculated parameters of [CI] line fits:} 
\section{Individual results for directly fitted and modeled [\ci] line profiles}
\input{resulttable}
%\clearpage
%\section{Other offsets in the [CI] line from the star:}\label{appendix-figs}

\begin{figure*}%[!!htbp]
\centering
\includegraphics[width=0.45\textwidth]{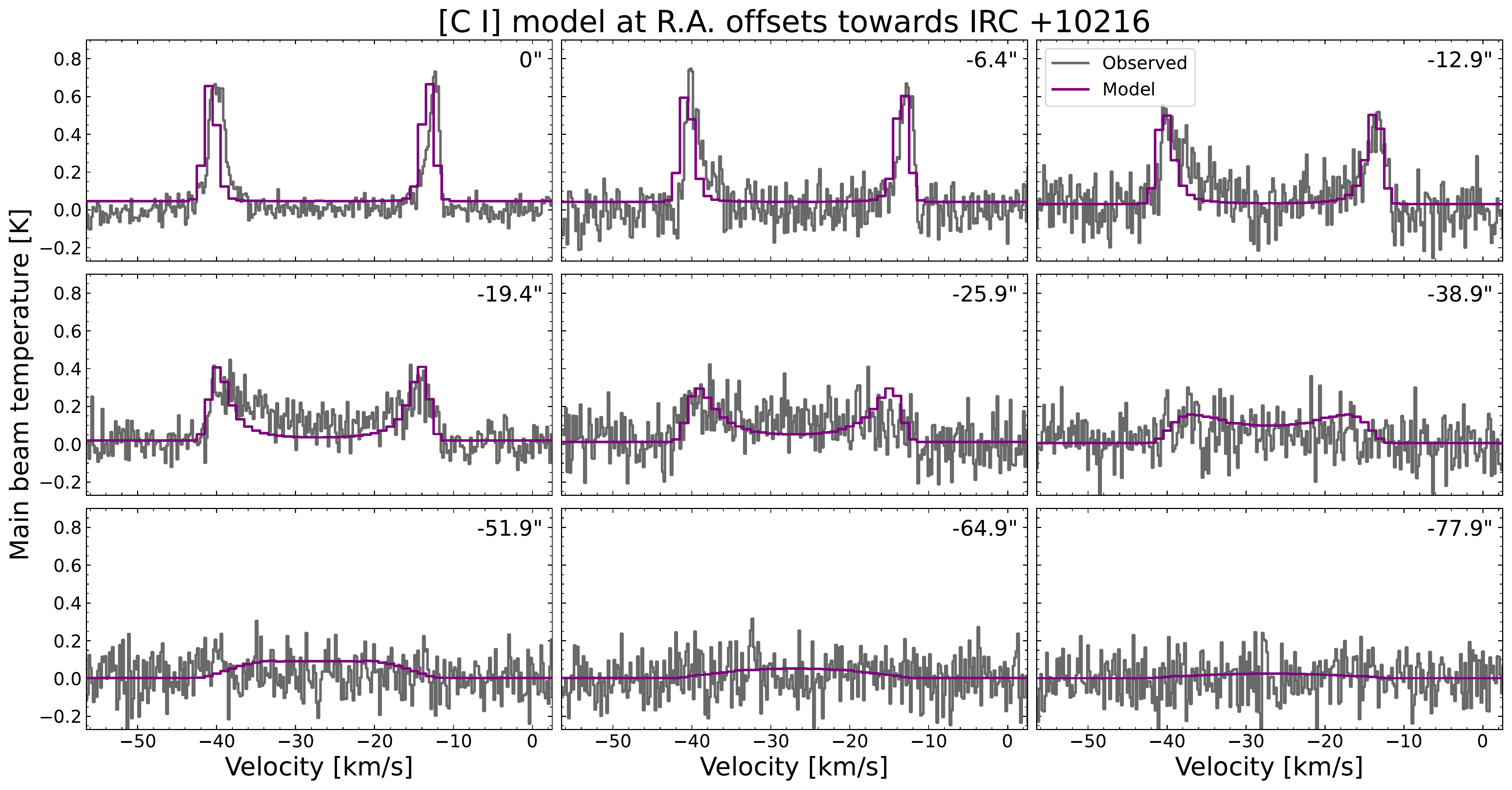}
%\caption{Plot showing observed [\ci] spectra \textit{(in black)} and the fit model \textit{(in purple)} at negative right-ascension offsets from the central position.}
%\label{fig:RAneg}
%\end{figure*}
%
%\begin{figure*}%![!htbp]
%\centering
\includegraphics[width=0.45\textwidth]{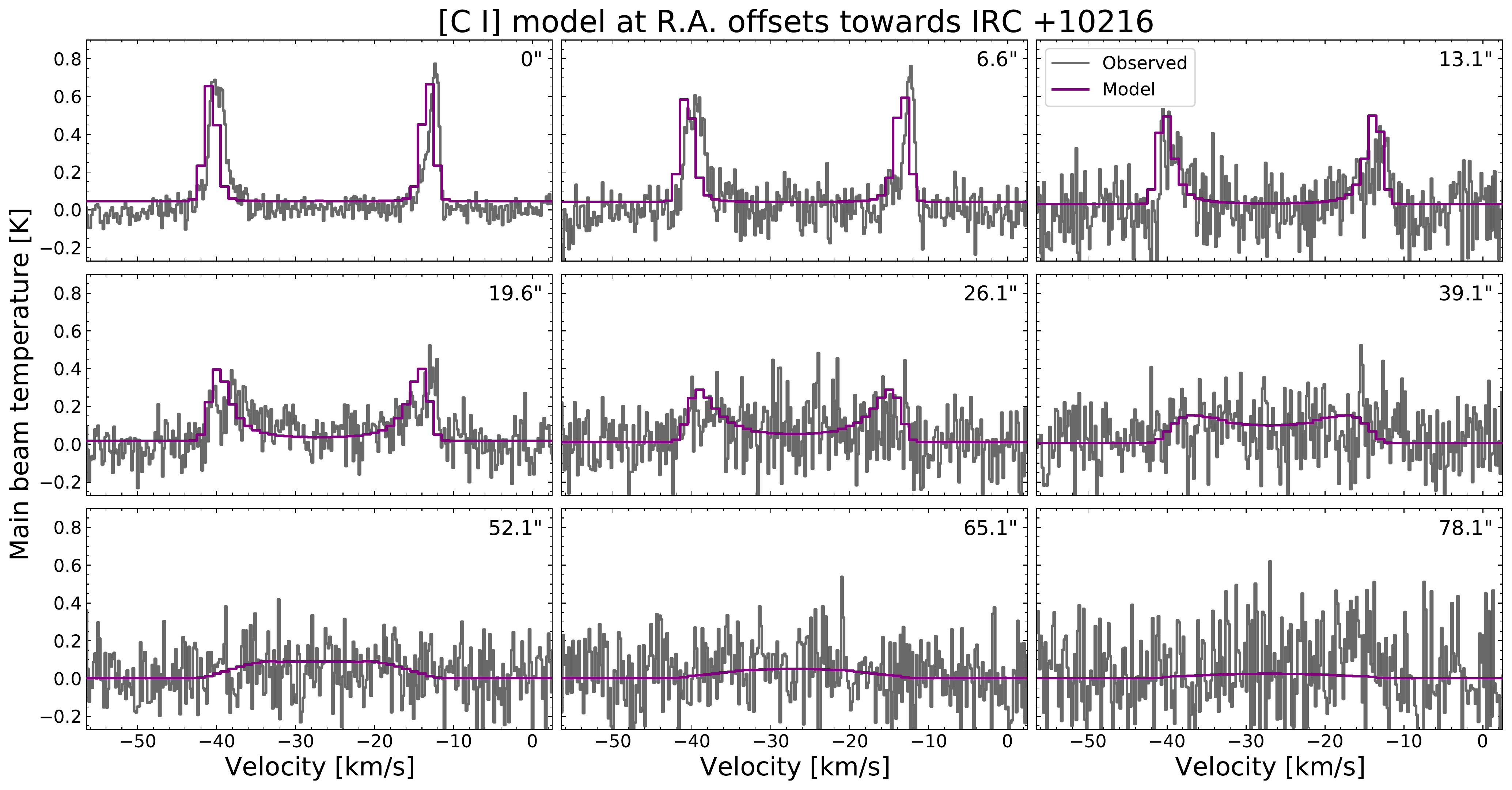}
%\caption{Plot showing observed [\ci] spectra \textit{(in black)} and the fit model \textit{(in purple)} at positive right-ascension offsets from the central position.}
%\label{fig:RApos}
%\end{figure*}
%
%
%\begin{figure*}%[!htbp]
%\centering
\includegraphics[width=0.45\textwidth]{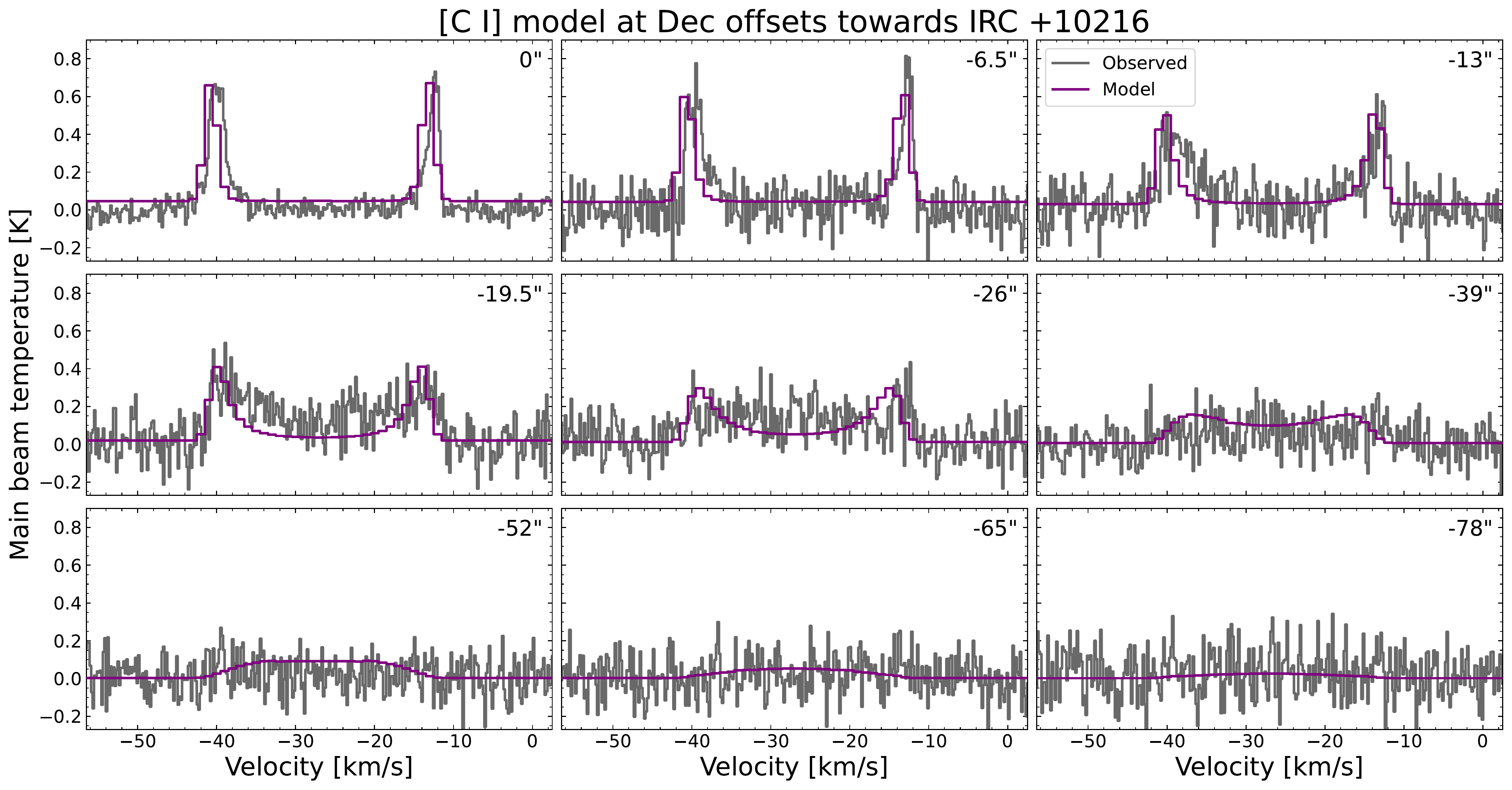}
%\caption{Plot showing observed [\ci] spectra \textit{(in black)} and the fit model \textit{(in purple)} at negative declination offsets from the central position.}
%\label{fig:Decneg}
%\end{figure*}
%
%
%\begin{figure*}%[!htbp]
%\centering
\includegraphics[width=0.45\textwidth]{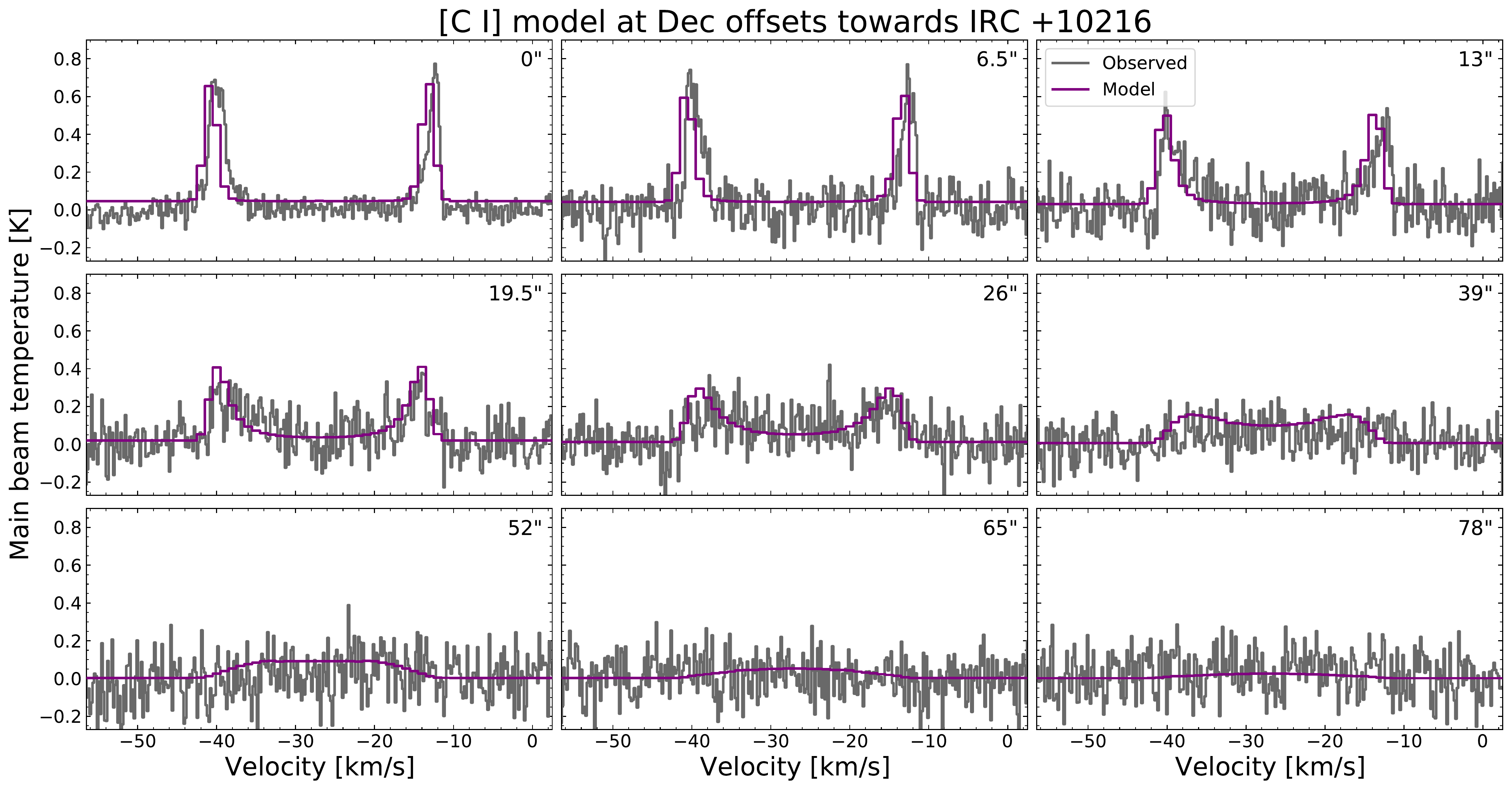}
\caption{Observed [\ci] spectra \textit{(in black)} and line profiles modeled with RATRAN \textit{(in purple)} at %positive declination offsets from the star.}
the sampled positions.}
\label{fig:IndividualPositionsModeled}
\end{figure*}

\begin{figure*}
    \centering
    \includegraphics[width=0.45\textwidth]{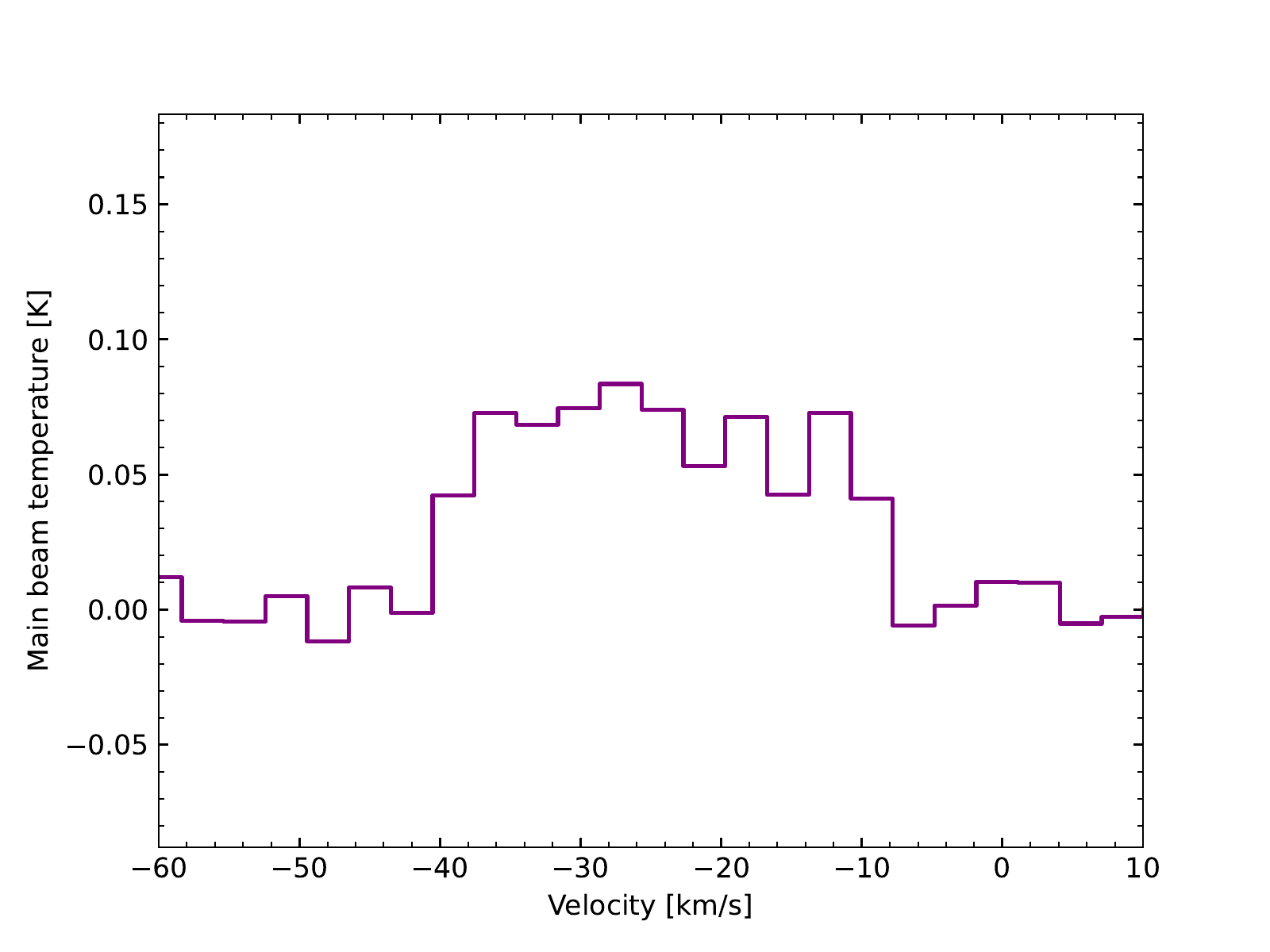}
    \caption{[\ci]~emission smoothed to 3~\kms at $39\arcsec$ from the star.}
    \label{fig:spec39offset}
\end{figure*}

\end{appendix}
\end{document}

%% file: RATRANmodel.tex
\begin{table}[!htb]
\scriptsize
\centering
\caption{\textbf{Input parameters for the RATRAN model of the [CI] emission}}
\begin{tabular}{cccccccc}
\hline
\hline
 D   & $\dot{M}$                    & $r_\text{in}$ & $r_\text{out}$    & $n_\mathrm{H_2}(r_\mathrm{in})$ & $T(r_\mathrm{in})$  & $\upsilon_\text{exp}$ & $\Delta\upsilon$ \\
$\mathrm{[pc]}$ & [M$_\odot$ $\text{yr}^{-1}$] & \multicolumn{2}{c}{\hrulefill\,[cm]\,\hrulefill} & $[\mathrm{cm^{-3}}]$ & [K] &   
 \multicolumn{2}{c}{[\kms]}   \\
\hline
% 130 & $2 \times {10^{-5}}$ & $3.9 \times {10^{14}}$ & $3 \times {10^{15}}$ & $5 \times {10^{-5}}$ & 14 & 1.5 \\ 
 130 & $2 \times {10^{-5}}$ & $2.3 \times {10^{16}}$ & $1.9 \times {10^{18}}$ & $4.5\times 10^4$ & 156 & 14.0 & 1.0 \\ 
\hline
\end{tabular}
\tablefoot{The [\ci] abundance profile is taken from \citep[][their Fig.~3]{Reach2022}. A second model uses a constant abundance (with respect to hydrogen nuclei) of $n(\mathrm{C^0})/n(\mathrm{H}) = 4\times 10^{-5}$ and $r_\text{in} =$$4 \times {10^{16}}$~cm. $D =$ distance to the source, $r_\text{in}$ and $r_\text{out} =$ inner and outer radius, respectively, of the shell, $\upsilon_\text{exp} =$ expansion velocity, and $\Delta\upsilon =$ line width (1/e half-width). The power-law indices for the fall-off of density and kinetic temperature (gas and dust temperature are assumed to be equal) with distance from the star are $-2$ and $-0.50$, respectively. For comparison, the temperature power-law index for purely adiabatic cooling at constant expansion velocity would be $-4/3$ \citep{CrosasandMenten1997}.}
\label{inputRATRANmodelparam}
\end{table}

%% file: resulttable.tex
\begin{table*}[ht!]
%\scriptsize
\centering
\caption{\textbf{Parameters deduced from direct fits to the observed spectra}}\label{resulttable}
\begin{tabular}{ccccccccc}
\hline
\hline
Offset & $V_1$ & $\int T_\text{MB} {\rm d}V_1$ & $\Delta V_1$& T$_\text{peak1}$& $V_2$ & $\int T_\text{MB}{\rm d}V_2$ & $\Delta V_2$ & T$_\text{peak2}$ \\

         & (\kms) &(K \kms)  &(\kms)& (K) &(\kms) & (K \kms)  & (\kms)  & (K)\\
\hline
 0\arcsec & $-$40 (0.03) & 1.8 (0.04) &2.5 (0.07) &0.67& $-$12.2 (0.08) & 1.3 (0.1)&1.7 (0.2)& 0.69 \\
 %\hline
&&&& \textbf{R. A. (negative)}\\
 %\hline
 $-$6.4\arcsec & $-$39.9 (0.1)&1.5 (0.15)&2.17 (0.2)&0.64&$-$12.9 (0.05)&1.3 (0.01)&1.9 (0.1)&0.66\\
 $-$12.9\arcsec & $-$38.8 (0.4) &2.3 (0.3)&5.9 (0.9)&0.36&$-$13.8 (0.2)&1.4 (0.1)&3.2 (0.4)& 0.43\\
%\textcolor{magenta}{$-$19.4\arcsec} & $-$35.9 (0.3)&3.15 (0.2)&12.6 (0.9)&0.23&$-$15.9 (0.3)&1.98 (0.1)&7.8 (0.9)&0.24\\
$-$19.4\arcsec & $-$36.3 (0.3) & 2.9 (0.2) & 11.2 (0.7) & 0.25 & $-$16.2 (0.3) & 2.07 (0.1) & 8.4 (1)& 0.23\\
% \rowcolor{lightgray} $-$25.9\arcsec & $-$29 (0.2) &3.74 (0.2)& 13.38 (0.9) & 0.13&&&&\\
$-$25.9\arcsec & $-$34.5 (1) & 2.9 (0.3) & 15.6 (2.1) & 0.17 & $-$20.3 (1.1) & 1.9 (0.3) & 12.4 (2.3) & 0.15 \\
%\rowcolor{lightgray} $-$38.9\arcsec & $-$29.8 (0.2) & 1.6 (0.2) & 10.8 (0.2) & 0.06&&&&\\
%$-$38.9\arcsec & 
% $-$51.9\arcsec & $-$\\
% $-$64.9\arcsec & $-$\\
% $-$77.9\arcsec & $-$\\
%\hline
&&&&\textbf{R. A. (positive)}\\
%\hline
6.6\arcsec & $-$39.6 (0.1)&1.67 (0.1)& 2.7 (0.2)& 0.57 &$-$12.5 (0.1)&1.06 (0.1)&1.35 (0.2)&0.74\\
13.1\arcsec & $-$38.95 (0.2)&1.15 (0.1)&3.4 (0.3)&0.315& $-$12.92 (0.2)&0.85 (0.1)&2.19 (0.4)&0.36\\
19.6\arcsec & $-$36.9 (0.5)&1.97 (0.2)& 9.02 (1.2)&0.21 &$-$14.25 (0.5)&1.6 (0.3)&6.6 (1.5)&0.23\\
26.1\arcsec & $-$34.1 (0.9) & 1.8 (0.2) & 14.1 (2.1)&0.11&$-$18.5 (1.4)&1.43 (0.3) & 13.2 (2.9) & 1.01 \\
%\rowcolor{lightgray} 26.1\arcsec & $-$27.2 (0.9) & 5.8 (0.4) & 14.6 (0.9) & 0.18&&&&\\
%\rowcolor{lightgray}39.1\arcsec & $-$28.4 (2.7) & 2.4 (0.2) & 19.06 (3.7) & 0.08&&&&\\
%\rowcolor{lightgray}52.1\arcsec & $-$30.7 (1.6) & 1.2 (0.2) & 10.52 (0.6) & 0.04&&&&check\\
%65.1\arcsec & $-$\\
%78.1\arcsec & $-$\\
 %\hline
&&&& \textbf{Dec (negative)}\\
 %\hline
 $-$6.5\arcsec&$-$39.6 (0.01)& 1.74 (0.1) &2.7 (0.2) &0.6 &$-$12.8 (0.01) & 1.39 (0.1)& 1.7 (0.2) & 0.72\\
 $-$13\arcsec& $-$38.8 (0.2) & 1.95 (0.1) & 5.5 (0.6) & 0.35 & $-$13.2 (0.1) & 1.13 (0.1) & 2.3 (0.2) & 0.47\\
$-$19.5\arcsec&$-$36.6 (0.3)&2.95 (0.2)&10.1 (0.8)&0.27 &$-$17.8 (0.5)&3.05 (0.2)&12.9 (1.1)&0.22\\
$-$26\arcsec & $-$32.5 (1.3)&3.02 (0.4)&17.2 (2.8)&0.17 & $-$20.7 (1.4) & 2.7 (0.4)&16.9 (2.7)&0.15 \\
%\rowcolor{lightgray} $-$26\arcsec & $-$27.6 (0.3) & 4.42 (0.2) & 16.1 (0.9) & 0.14&&&&\\
%\rowcolor{lightgray} $-$39\arcsec & $-$24.7 (2.5) & 2.09 (0.2) & 17.7 (2.2) & 0.07&&&&\\
% $-$52\arcsec\\
% $-$65\arcsec\\
% $-$78\arcsec\\
 %\hline
&&&&\textbf{Dec (positive)}\\
 %\hline
 6.5\arcsec & $-$39 (0.1) & 1.6 (0.1) &2.3 (0.2) & 0.6 &$-$12.7 (0.1)& 1.2 (0.01)& 1.7 (0.1)&0.66\\
 13\arcsec & $-$39.3 (0.3) & 1.4 (0.2) & 3.3 (0.6) & 0.38 & $-$13 (0.1) &  1.3 (0.1) & 3.0 (0.4) & 0.39\\
19.5\arcsec& $-$37.53 (0.4)& 1.83 (0.2)&8.3 (1)&0.21&$-$14.5 (0.3)&1.14 (0.1)&4.10 (0.7)& 0.26\\
26\arcsec &$-$34.8 (0.8)&1.7 (0.2)& 11.0 (1.6) & 0.15 & $-$17.4 (0.8)&1.9 (0.2) & 11.9 (1.7)& 0.15\\ 
%\rowcolor{lightgray} 26\arcsec & $-$26.4 (0.8) & 3.38 (0.2) & 14.8 (0.8) & 0.09&&&&\\
%\rowcolor{lightgray} 39\arcsec & $-$25.9 (0.2) & 1.78 (0.2) & 15.06 (0.2) & 0.06&&&&\\
% 52\arcsec\\
% 65\arcsec\\
% 78\arcsec\\
\hline
\end{tabular}
\tablefoot{The table summarises physical parameters constrained at different offsets from the star. Subset 1 refers to the blue-shifted peak and subset 2 refers to the red-shifted peak. The parameters obtained for offsets  away from the central position are calculated by fitting the Gauss method. $V$ is the LSR velocity, $\int T_\text{MB} {\rm d}V$ is the calculated integrated intensity, $\Delta V$ refers to the line width calculated from the Gauss method, and finally T$_\text{peak}$ is the peak temperature of the fitted line.}
\end{table*}
%\normalsize